# Strain induced lithium functionalized graphane as a high capacity hydrogen storage material

Tanveer. Hussain,[1] Abir. De Sarkar,[1,2] and Rajeev. Ahuja[1,2]

[1]Condensed Matter Theory Group, Department of Physics and Astronomy, Box 516, Uppsala University S-75120 Uppsala, Sweden, EU

[2]Applied Materials Physics, Department of Materials and Engineering, Royal Institute of Technology (KTH) S-100 44 Stockholm, Sweden, EU

Strain effects on the stability, electronic structure, and hydrogen storage capacity of lithium-doped *graphane* (CHLi) have been investigated by state-of-the art first principle density functional theory (DFT). Molecular dynamics (MD) simulations have confirmed the stability of Li on graphane sheet when it is subject to 10% of tensile strain. Under biaxial asymmetric strain, the binding energy of Li of *graphane* (CH) sheet increases by 52% with respect to its bulk's cohesive energy. With 25% doping concentration of Li on CH sheet, the gravimetric density of hydrogen storage is found to reach up to 12.12wt%. The adsorption energies of $H_2$ are found to be within the range of practical $H_2$ storage applications.

The ever-increasing world wide demand of energy and the global warming caused by the escalating $CO_2$ emission due to the rapid depletion of fossil fuels underscores the urgent need for alternative, clean and renewable sources of energy. Hydrogen ($H_2$) can potentially serve as an alternative energy carrier because it is safe, cheap and abundant in nature. Furthermore, it is environmentally friendly and has the highest energy density than any other element.[1,2] However, the main obstacle in the realization of $H_2$ economy and in the practical implementations of $H_2$ as an efficient energy carrier is its gaseous nature and unavailability of feasible storage media. Some of the common techniques employed in $H_2$ storage include high-pressure tanks, low temperature liquid tanks, metal hydrides as storage materials, carbon based storage materials etc.[3] In most of the storage materials low storage capacity, high operational temperature, absence or low reversibility, slow kinetics and safety issues preclude their practical applications.

Carbon based nanostructures (carbon nanotubes, fullerenes, graphene etc.) are considered to be the excellent storage materials of $H_2$ because of their lightweight and low cost. They can absorb/desorb high capacity of $H_2$ at ambient conditions. Carbon nanostructures have been explored for $H_2$ storage in several theoretical as well as experimental studies.[4-9] In pristine form; the carbon nanostructures are usually very inert and bind gas molecules very weakly. This inertness shows up as a main hindrance in the actualization of its several possible technological applications. So, they are functionalized by dopants for the purpose of enhancing their binding or adsorption capabilities.

Metal functionalized carbon nanostructures have immense potential for a diverse array of applications such as molecular sensors, biomedicine, nanotechnology, catalysis, energy storage and so on. Regardless of the purpose of functionalization, it needs to meet these two criteria simultaneously, (i) stability and (ii) efficiency. First of all, the dopant atom should bind stably to the nanomaterial. Then, the functionalization needs to serve its intended purpose optimally. Several avenues are being explored to meet the two aforementioned criteria for metal functionalization.

Light metals are preferred to the transition metals for the purpose of functionalization as the later shows a high propensity for clustering due to their high cohesive energy, which in turn lowers the storage capacity of $H_2$.

Graphene, which is a single atomic layer of graphite possessing extraordinary electrical, mechanical, and optical properties, has stimulated a surge of research interests right after its experimental isolation.[10] Along with countless other technological applications, graphene doped with light metals has drawn considerable attention for its potential applications in $H_2$ storage. Recently Beheshti et al.[11] has reported calcium-decorated graphene as promising material for hydrogen storage. Their density functional theory study reveals that boron substitution on graphene circumvents the clustering of doped Ca atoms and Ca doped on both sides of graphene helps to attain a storage capacity of 8.38 wt%. Park et al.[12] have investigated the lithium functionalized graphene system for high capacity $H_2$ storage material. They also mentioned that the Li-graphene binding energy increase significantly with boron doping. The storage capacity of $H_2$ has been reported to reach up to 13.2 wt%. Besides light metal adatoms, polylithiated structures such as $OLi_n$ and $CLi_m$ (n=1-4, and m=3-5) have also been decorated on graphene layer for the purpose of $H_2$ storage.[13] Graphene has been demonstrated to store a reasonably good amount of $H_2$ (~8.5wt %) when it is functionalized by these structures.

An interesting structure called *graphane (CH)* or hydrogenated graphene is formed when a pure graphene sheet is exposed to hydrogen plasma. The hydrogen atoms attach covalently to each carbon atom alternately on both sides of the graphene sheet resulting into a crumbled structure in contrast to a planar graphene. It was theoretically predicted by Sofo et al.[14] and experimentally verified by Elias et al.[15]

Due its nano size, structural stability and its ability to bind metals strongly, *graphane* has been investigated as a potential $H_2$ storage material.[16,17]

For an optimal utility of functionalization, a uniform distribution of metal adatoms on the carbon-based substrate (graphene, *graphane*) is necessary in order to avoid clustering between metal adatoms and to store $H_2$ efficiently. Beside this the binding of $H_2$ molecules to the host material (metal-graphene or metal-*graphane*) will determine the feasibility of sorption and desorption of $H_2$ at ambient conditions and the reversibility between the two processes. A weak binding will lower the storage capacity, while a strong binding will lessen the probability of $H_2$ release.[18]

Different approaches have been employed to counteract the clustering proclivity of the metal dopant atoms on the surface of carbon nanostructures and to bind $H_2$ strong enough for practical applications. One of the most extensively used technique is boron substitution.[11,12,19,20]

Zhou et al.[21] have demonstrated the application of electric field to enhance the binding $H_2$ to polarizable substrate like boron nitride. Shevlin et al.[22] reported the improvement in the binding of Ti to carbon nanotubes by creating vacancy. Despite all these reported studies, stabilizing the functionalization of metal atoms on carbon nanostructures still continues to be an ongoing issue, which needs to be addressed very carefully.

This study deals with the strain-induced stabilization of lithium adatoms on *graphane* sheet under the application of mechanical strain. Two different symmetries in the dispersion of Li atoms on *graphane* namely hexagonal and rectangular have been considered to compare their relative stability. The effects of Li stabilization on the enhancement in the $H_2$ storage capacity of lithium-doped *graphane* (CHLi) system have also been discussed. More than 15% strain has been easily realized experimentally in graphene and graphene is found to lie within its elastic limits under this magnitude of strain.[23,24] A couple of years ago Topsakal et al. [25] applied strain on *graphane* to study its mechanical properties. They have also studied the variation of band gap with applied strain. There have been no studies on strain application in metal (lithium) doped *graphane* for the purpose of enhancing its hydrogen storage efficiency. To that end, we have performed state-of-the-art first principle calculations using density functional theory to probe into the strain-induced stabilization of Li functionalized *graphane* and its impact on $H_2$ storage.

In this study the total energy and electronic structures calculations have been carried out by means of projector augmented wave method of density functional theory (DFT) as implemented in the Vienna *ab initio simulation package* (VASP). [26] Both local density approximation (LDA) and the Perdew-Burke-Ernzerhof (PBE) variant of the generalized gradient approximation (GGA) has been employed for the exchange and correlation potential.[27,28]

The Molecular dynamics runs have also been performed to examine the stability of Li atom on *graphane* sheet under strain by using projector-augmented wave approach[29] as implemented in VASP.

Recently, Cha.et al.[30] showed that the adsorption of $H_2$ calculated by DFT on calcium (Ca) cations centers calculated by DFT results in binding energies, which are unphysical if compared with the corrected wave function theories. In response to this Ohl. et al. [31] were of the opinion that the inadequate description of the polarization functions in 6-311++G** basis set overestimates the binding energies mentioned by Cha. et al.

The short-range correlation effects will be accounted for and the binding energies of DFT will be comparable to that of MP2 if large basis sets are used.

Taking these facts into consideration, we have calculated the van der Waals corrected interaction energies using the semiempirical correction of Grimme[32] as implemented in VASP. This will allow a better treatment of the weakly interacting systems.

The unit cell of graphane consists of four atoms (2C, 2H). A 2x2x1 super cell containing 8 carbon and 8 hydrogen atoms has been used in this study. A vacuum thickness of 15Å in the (001) direction, perpendicular to the *graphane* sheet, has been used to decouple the periodic images of *graphane* sheet. The Brillouin zone has been sampled using Monkhorst-Pack 10x10x1 K-mesh. For obtaining the density of states (DOS) we have used 17x17x1 k-mesh. All the structures were fully relaxed until the force acting on each ion reached less than 0.005 eV/Å.

Before discussing the effects of strain on the binding energy or stabilization of Li adatom on *graphane*, the geometry of pure *graphane* is described briefly. FIG.1(a, b) shows the side and top view of the optimized structure of pure *graphane.* In the fully optimized relaxed structures, the C-C and C-H bond lengths are 1.53 Å and 1.12 Å respectively, which are in good agreement with the previous study.[14] The effect of Li doping on *graphane* has been examined by substituting two out of the eight hydrogen atoms (one from +Z and one from –Z direction) with Li atoms resulting in a doping concentration of 25%.

FIG. 1 (c, d) shows the side and top view of the optimized structures of CHLi. The CH-Li bond length is found to be 2.08 Å upon optimization.

Binding energy ($\Delta E_b$) which can be defied as "*the amount of energy required to take the adatoms (Li in this case) away from graphane sheet such that its interaction with*

*the sheet becomes negligible*" is the most important factor in realizing a uniform distribution of adatoms on *graphane* surface. The higher the value of $\Delta E_b$, the lesser is the possibility for clustering among Li adatoms and higher the chances of having a uniform dispersion of adatoms on the *graphane* surface. $\Delta E_b$ has been calculated by the relation

$$\Delta E_b = E(CHLi) - E(CH) - E(Li)$$

Where E (CHLi), E (CH), and E (Li) are the total energies of $C_8H_6Li_2$, $C_8H_6$ and isolated Li atom respectively. It should be noted that putting it at the center of a cubic super cell of length 20 Å and employing Γ-point sampling of the Brillouin zone calculate the energy of isolated Li atom. Atomic geometries of all other structures were fully relaxed, while keeping the volume fixed and allowing variations in shape. As mentioned earlier, two different symmetries in the dispersion of Li adatoms on *graphane* sheet namely hexagonal and rectangular have been considered in this study. The values of $\Delta E_b$ in case of hexagonal and rectangular structures without strain are 1.755 eV and 2.231 eV respectively, implying that the dispersion of Li atoms in the rectangular symmetry on graphane sheet is energetically favorable. To avoid underestimation and overestimation caused by GGA and LDA respectively, we have also employed van der Waal's corrected dispersion term in the GGA-PBE for the calculation of $\Delta E_b$.

Then, the variation in the binding energies with the extent and symmetry of mechanical strain (Uniaxial, Biaxial, Symmetric, Asymmetric) for both the Li dispersion symmetries in the CHLi are discussed and the results are tabulated in the FIG.2 and FIG. 3.

In case of hexagonal cell, a significant strain-induced increase in the binding energy of Li on *graphane* sheet is evident from FIG. 2. The maximum value of $\Delta E_b$ (2.401 eV) can be obtained with a biaxial strain (X=7.5%, Y=10%), which is 38% higher than the $\Delta E_b$ without strain. In case of rectangular symmetry, the increase in $\Delta E_b$ with strain is less pronounced, as indicated in FIG. 3, but under a biaxial strain of (X=7.5%, Y=10%) a maximum in $\Delta E_b$ (2.48 eV) is found to be attained, which is almost 52% higher than the cohesive energy of Li (1.63 eV) [20] in bulk. Our results show that there is a significant increase in $\Delta E_b$ in case of CHLi under mechanical strain. The higher value of $\Delta E_b$ of Li on the sheet (2.48 eV in case of rectangular cell) is an indication of the stability of CHLi monolayer at higher temperature and higher doping concentration.

The stability of Li atoms on graphane sheet has been further confirmed by applying molecular dynamics simulations. The Nose-thermostat algorithm at 400 K with 1fs time step has been used. The structure of CHLi has been found to remain intact after 3 ps, ensuring the stability of the structure, also shown in FIG. 4.

To examine the type of Li-CH bonding, we have performed the Bader charge analysis [33]. The electronegativity of C is much higher than that of Li. So, the electronic charge is transferred from Li to C in *graphane* sheet. After analyzing, it has been found that 0.981 electronic charge has been transferred from each of the Li atoms on either side to the sheet. Most of this charge from Li is transferred to C atoms to which Li atoms are covalently bonded. Some of the charge also goes to the H atoms, which are in the close vicinity. The C atom which is directly attached to Li on either side of the sheet gets ~0.607 electronic charge and the nearby H atoms attains 0.169 e. Due to this charge transfer, Li gains a fractional positive charge, whereas C and H gather a fractional negative charge, which is clear from iso-surface of differential charge density shown in FIG. 5.

Furthermore, the total and partial density of states of pure as well as Li doped

*graphane* has been plotted and analyzed. In case of pure *graphane*, as evident from FIG. 6, an insulating behavior with a wide band gap of 3.55 eV is predicted. This value is in good agreement with the previous study [14]. In case of CHLi, as shown in the total and the partial density of states in FIG. 6, a drastic change in the band gap is found to occur, which results in a metallic behavior of system CHLi. This insulator to metallic transition is induced by the states that show up in the original band gap of the pure CH. If the p DOS of CHLi is closely examined, it reveals that the states appearing just above the Fermi level are due to the strong hybridization among the Li atom, the C atom to which Li is covalently attached and the H atom which lies in the close vicinity of Li atom. The band, which starts at the Fermi level ($E_F$) and goes further beyond the $E_F$ is solely contributed by Li. The other C and H atoms have not contributed in this region, indicating that the interaction of Li with CH is localized.

Finally, we discuss the adsorption of $H_2$ molecules on CHLi system. For this purpose we have considered the rectangular cell of CHLi. First, we have considered the adsorption of $H_2$ molecules on CHLi without strain. The $\Delta E_b$ of Li in this case is 2.2 eV, which is clearly lower than that of $\Delta E_b$ (2.48 eV) under strain. The $H_2$ molecules are introduced on the fractionally positively charged $Li^+$ ions on the both sides of the sheet in a stepwise manner. The mechanism of the adsorption of $H_2$ on CHLi could be explained by considering the fact that positively charged $Li^+$ ions polarizes the $H_2$ molecules, which are then held to the $Li^+$ ions by the van der Waal's attractive forces. In order to maximize the storage capacity, the $H_2$ molecules bind themselves to the Li ions at physisorption distances while maintaining reasonable distances among themselves. This minimizes the repulsion among the adsorbed $H_2$ molecules. It has also been found that at the most $3H_2$ molecules can be adsorbed on each Li in unstrained CHLi system resulting in a storage capacity of 9.37 wt%. By applying biaxial asymmetric strain (X=7.5%, Y=10%) a stronger Li-CH binding is achieved. This not only increases the stability of Li and ensure the uniform distribution of Li on the CH sheet but also enhances its storage capacity. At the most $4H_2$ is found to be adsorbed on each Li atom on both sides of the sheet with reasonably good adsorption energy. FIG.7 shows the optimized geometry of CHLi at the maximum possible coverage of $H_2$ molecules. This results in a very high storage capacity of 12.12wt%, which goes well beyond the DOE target to be attained till 2017.

The following relation has been used to calculate $\Delta E$

$$\Delta E (n) = E\{(CHLi+nH_2) - E(CHLi) - E(H_2)\}/n$$

Where $\Delta E(n)$ is the adsorption energy of the nth $H_2$ molecule adsorbed on the CHLi sheet, E(CHLi) is the energy of CHLi sheet without $H_2$ molecule and $E(H_2)$ is the energy of a single $H_2$ molecule.

TABLE. 1 shows the complete results of adsorption energies $\Delta E$ (eV) of $H_2$ molecules adsorbed on CHLi and the H-H distance $\Delta d$ (Å).

To cope with the overestimation of LDA and underestimation of GGA and to bring reliability in the calculation of $\Delta E$ and H-H distance, we have incorporated the van der Waal's corrected dispersion term in our calculations. Table 1 shows a clear consistency in the values of adsorption energies of $H_2$ on CHLi and H-H bond lengths, regardless of the method of calculation.

Using first-principle calculations, we have demonstrated that Li-doped *graphane*

(CHLi) can be used a high capacity hydrogen storage material. The binding energy of Li on CH sheet has been calculated with and without strain. It is found that with a biaxial asymmetric strain (X=7.5%, Y=10%) the binding energy of Li can reach 52% higher than its bulk's cohesive energy. To ensure the stability and uniformity in the distribution of Li on CH molecular dynamics simulations has also been carried out. A reasonably high $H_2$ storage capacity of 12.12wt% has been obtained with 25% of Li doping on CH sheet. The adsorption energies of $H_2$ on CHLi are found to be within the ideal range of 0.15 eV- 0.20 eV suitable for the practical storage applications.

TH is thankful to higher education commission of Pakistan for doctoral fellowship. ADS is grateful to the Wenner-Gren Foundation for postdoctoral Fellowship. RA acknowledges FORMAS, SWECO and Wenner-Gren Foundation for financial support. SNIC and UPPMAX are acknowledged for computing time.

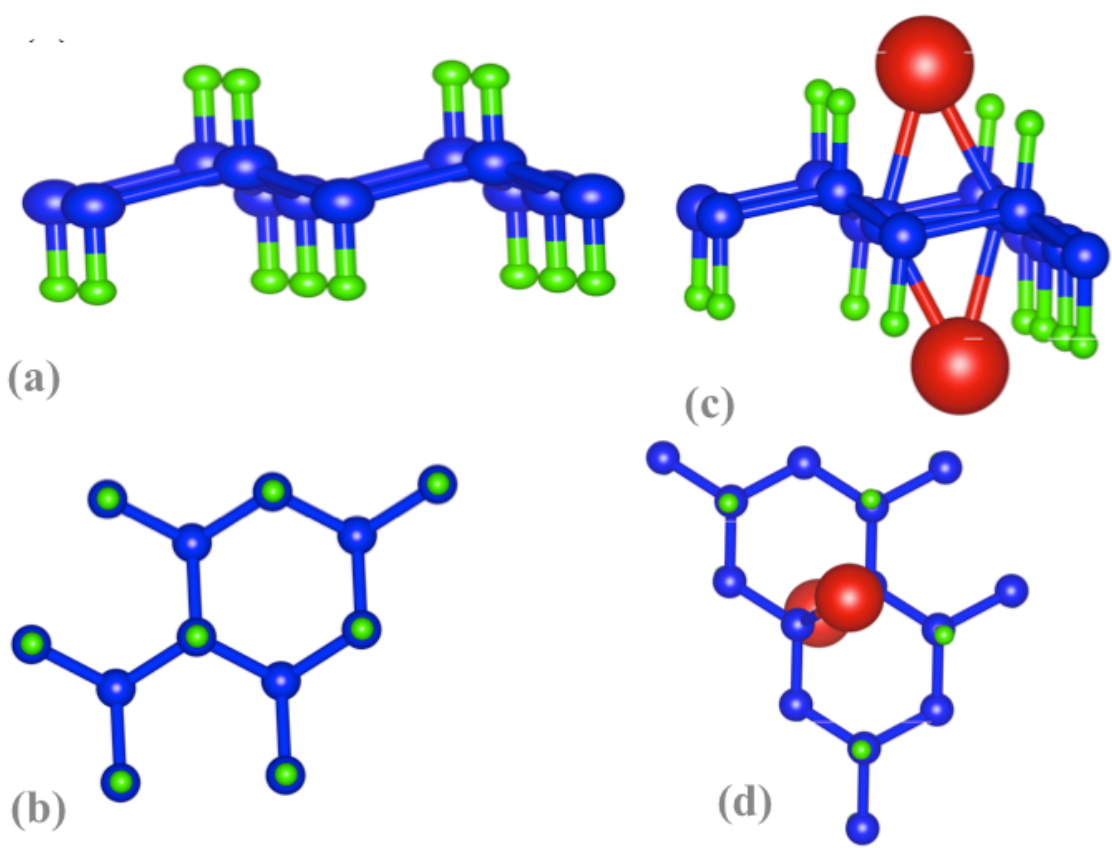

FIG.1. (a) and (b) Side and top views of the optimized structure of pure CH respectively, (c) and (d) Side and top views of the optimized structure of CHLi respectively.

FIG. 2. Binding energies of Li atom in the hexagonal CHLi cell under strain calculated by LDA, GGA and van der Waal's induced calculations.

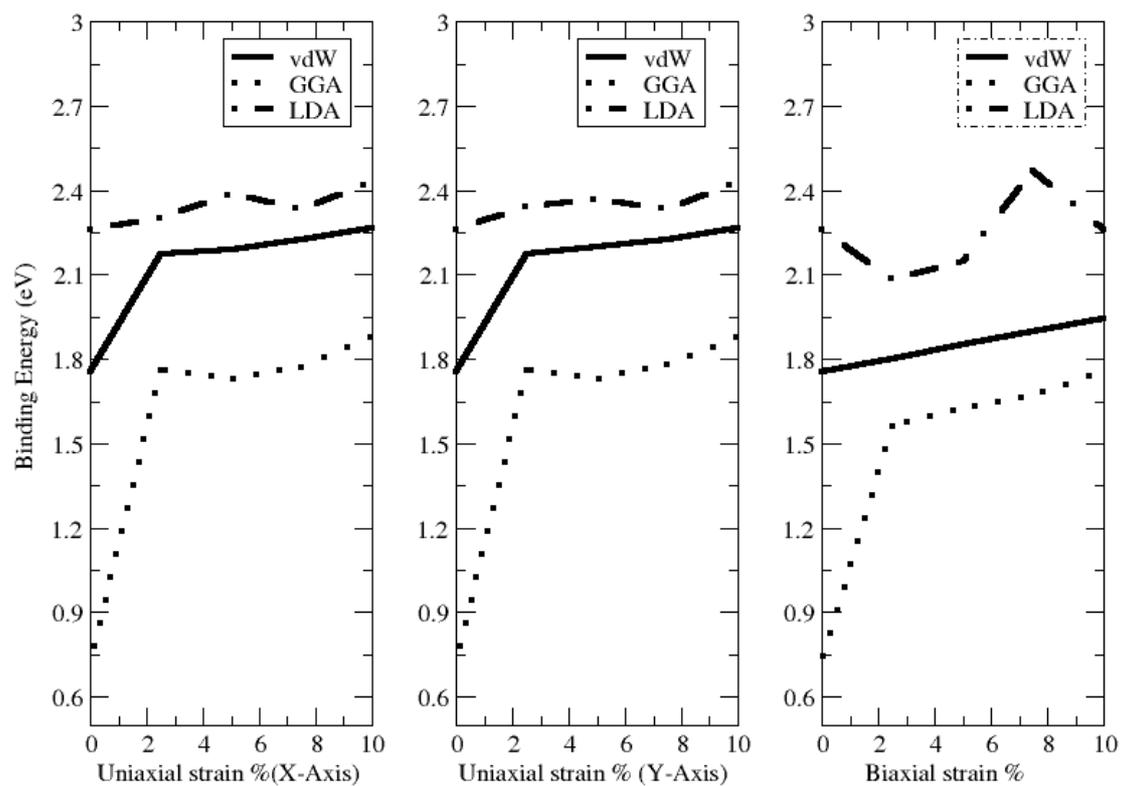

FIG. 3 Binding energies of Li atom in the rectangular CHLi cell under strain calculated by LDA, GGA and van der Waal's induced calculations.

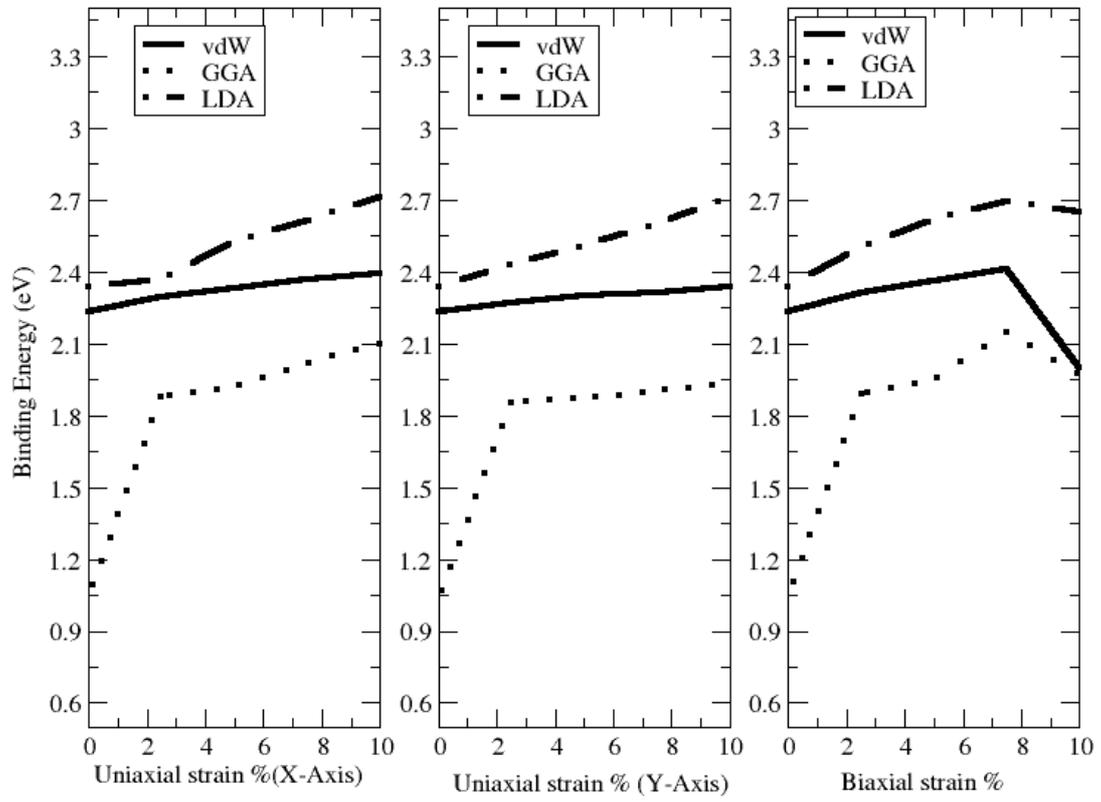

TABLE. 1. Adsorption energies ΔE (eV) of the $n^{th}$ (n=1-4) hydrogen molecule adsorbed on Li-doped *graphane* (CHLi) under asymmetric biaxial strain (X=7.5%, Y=10%), and average H-H bond distance Δd (Å) is calculated by using LDA, GGA and van der Waals interaction included Calculations.

| No. of $H_2$ Molecules | LDA | | GGA | | Van der Waal's | |
| --- | --- | --- | --- | --- | --- | --- |
| | ΔE(n) | Δd | ΔE(n) | Δd | ΔE(n) | Δd |
| 2 | 0.280 | 0.808 | 0.140 | 0.779 | 0.150 | 0.780 |
| 4 | 0.228 | 0.800 | 0.109 | 0.785 | 0.120 | 0.782 |
| 6 | 0.205 | 0.799 | 0.100 | 0.774 | 0.130 | 0.778 |
| 8 | 0.175 | 0.796 | 0.095 | 0.765 | 0.136 | 0.776 |

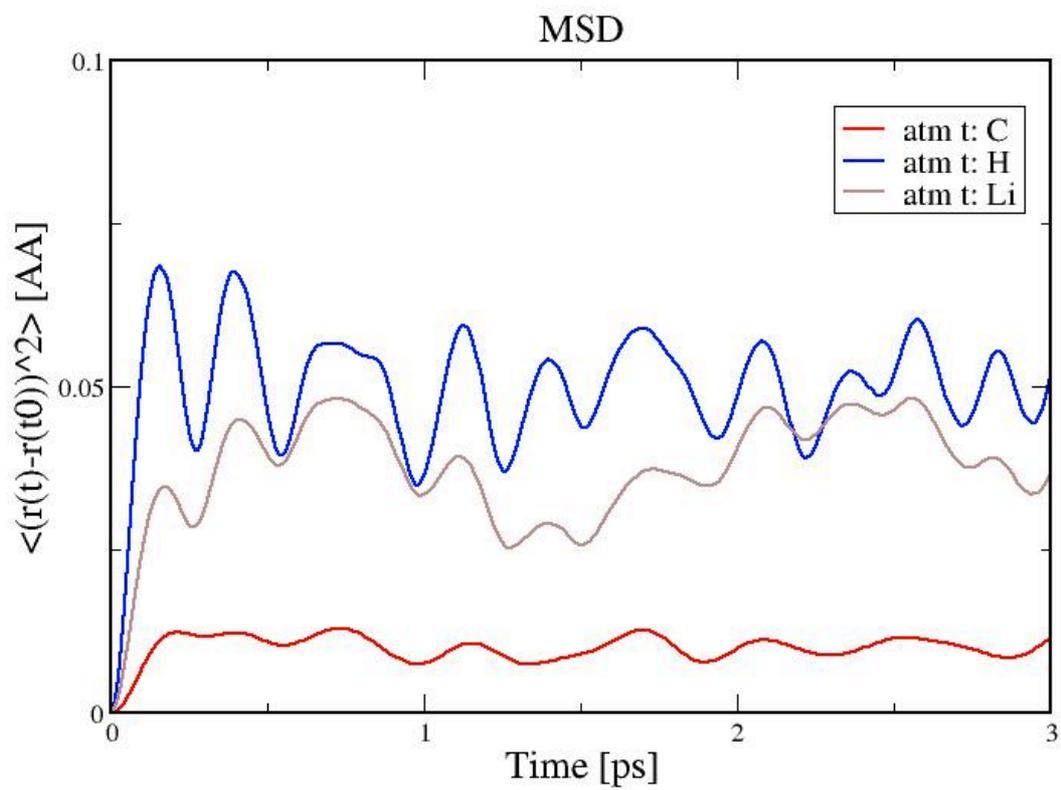

FIG. 4. Mean square displacement (msd) with time obtained from molecular dynamics simulation of CHLi sheet.

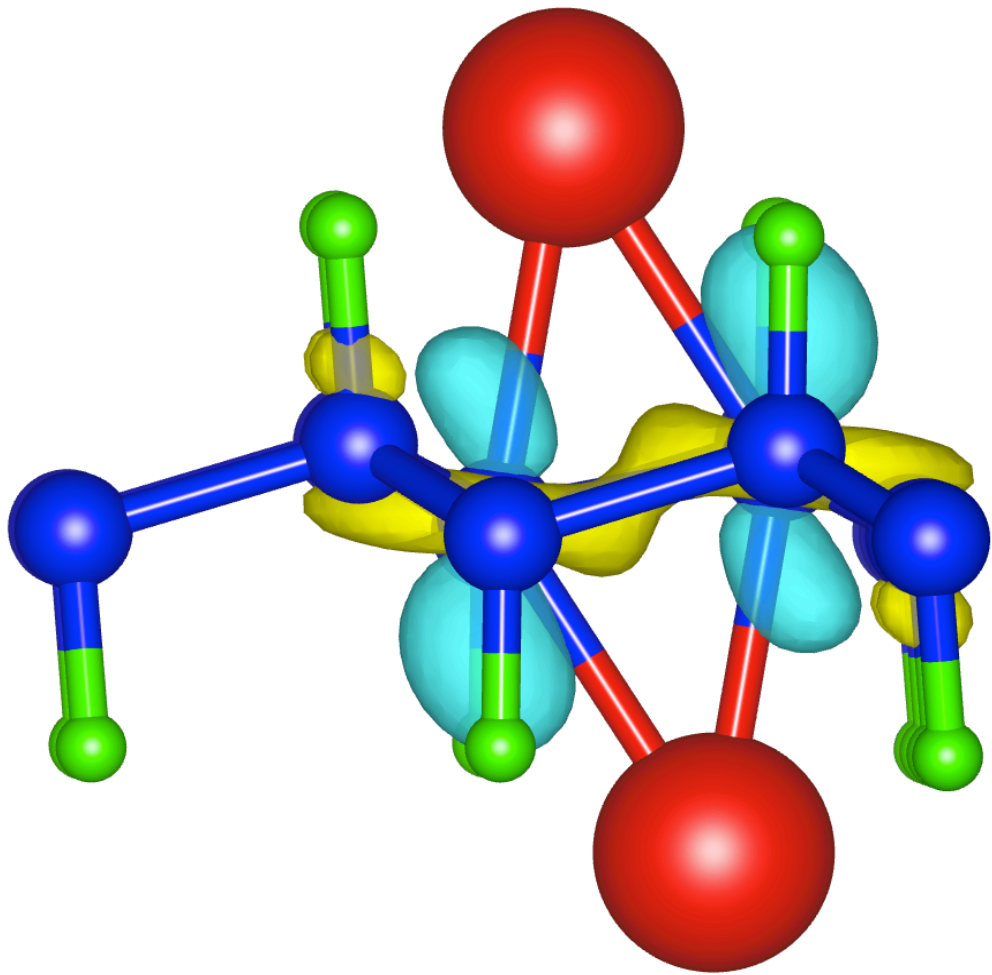

FIG. 5. Isosurface of the differential charge density calculated as

$$\Delta\rho = \rho\,(CHLi) - \rho\,(CH+Li)$$

Cyan and yellow colors indicate accumulation and depletion of charges respectively.

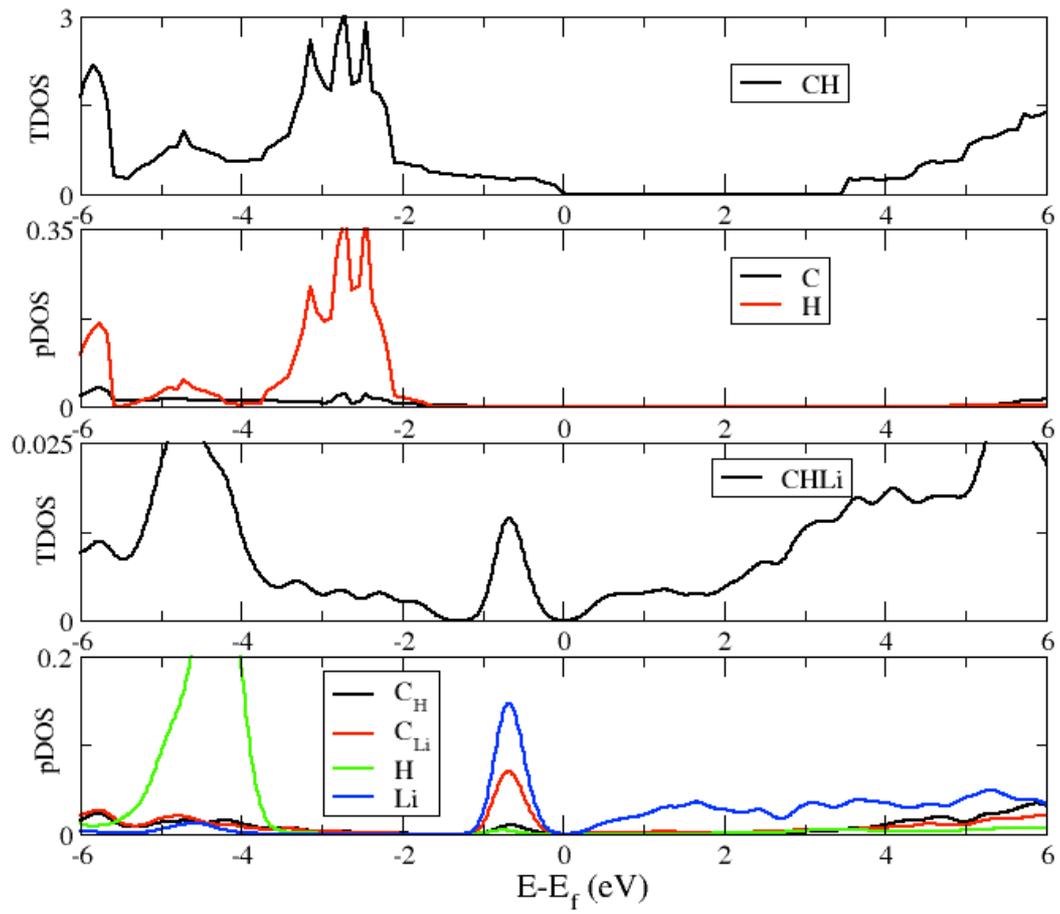

FIG. 6. Total and partial density of states of CH and CHLi. Fermi level is set to 0.

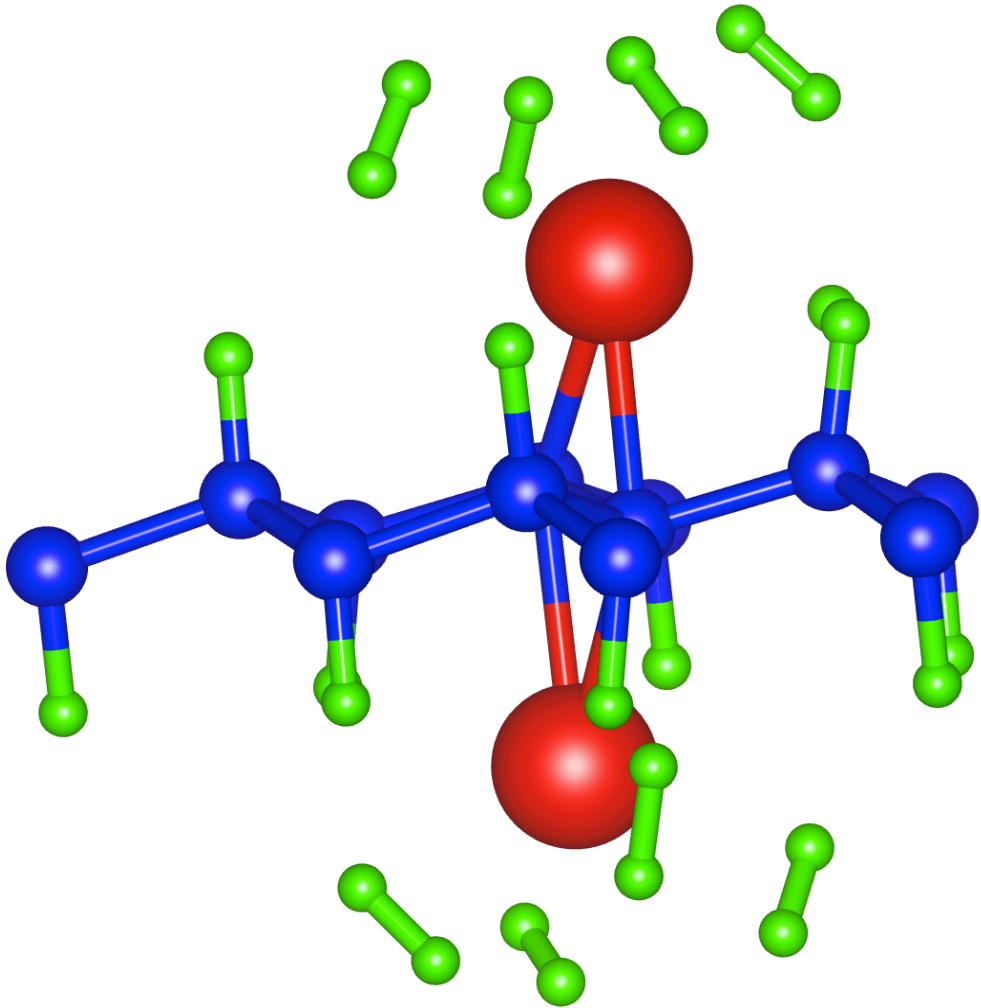

FIG. 7. Optimized structure of CHLi with maximum coverage of $H_2$ molecules under biaxial asymmetric strain (7.5%, 10%).